\documentclass[smallextended]{svjour3}

\smartqed 

\usepackage{graphicx,amssymb,amsmath}
\usepackage{subfig}

\begin{document}

\title{Modelling Inhomogeneity in Szekeres Spacetime}

\author{David Vrba  \and
Otakar Sv\'{\i}tek
}

\institute{Institute of Theoretical Physics, Faculty of Mathematics and Physics, Charles University in Prague, V~Hole\v{s}ovi\v{c}k\'ach 2, 180~00 Prague 8, Czech Republic \\
\email{vrba.david@centrum.cz, ota@matfyz.cz}
}
\date{Received: date / Accepted: date}

\maketitle

\begin{abstract}
We study the behaviour of the density contrast in quasi-spherical Szekeres spacetime and derive its analytical behaviour as a function of $t$ and $r$. We set up the inhomogeneity using initial data in the form of one extreme value of the density and the radial profile. We derive conditions for density extremes that are necessary for avoiding the shell crossing singularity and show that in the special case of a trivial curvature function, the conditions are preserved by evolution. We also show that in this special case if the initial inhomogeneity is small, the time evolution does not influence the density contrast, however its magnitude homogeneously decreases. 
\keywords{inhomogeneous cosmology \and Szekeres spacetime}
\PACS{04.20.Jb, 04.20.Dw}
\end{abstract}

\section{Introduction}
Homogeneous cosmological models have successfully explained many important features of our universe. However, we know that the distribution of matter is not homogeneous and so these models are only an approximation. In the last decade modelling of inhomogeneity in cosmology has become a topic of substantial interest. There are several different approaches to the problem. Due to the nonlinear nature of Einstein equations one should not rely on perturbation theory completely therefore exact models with nonuniform distribution of matter should be considered as well. Among the most studied exact inhomogeneous solutions of Einstein equations belong Lemaitre-Tolman-Bondi (LTB) metric \cite{1}, Szekeres solution \cite{2}, Szafron family of solutions \cite{3}, Stephani solution \cite{4} or Lemaitre metric \cite{1}, which is a generalization of LTB for a fluid with nonzero pressure. An overview of inhomogeneous cosmological solutions can be found in \cite{6}. 

LTB metric was extensively studied by Krasinski and Hellaby to model structure formation \cite{7} - \cite{9}. Together with Bolejko they used the solution to describe formation of voids in the universe \cite{10}. 
LTB metric is a special case of the Szekeres solution that was discovered by Szekeres \cite{2} and was developed by Hellaby and Krasinski in a series of papers \cite{11} - \cite{14}. They gave a geometrical interpretation of the metric functions in all three different types of geometry that are quasi-spherical, quasi-pseudospherical and quasi-planar. The quasi-spherical case is currently the best understood of all three types. It has found a cosmological application in the study of Bolejko who focused on structure formation \cite{15}, \cite{16} and constructed models of a void with an adjourning supercluster. Bolejko also investigated Buchert averaging of the quasispherical Szekeres metric \cite{17}, constructed a Szekeres-Swiss-cheese model that he used to estimate the impact of inhomogeneity on the propagation of light  \cite{18} and CMB observations \cite{19}.

Walters and Hellaby showed in a recent paper \cite{20} how to model inhomogeneity in quasispherical Szekeres metric using initial and final data. They constructed three models where they specified initial and final radial density profile and one extreme value of the density on the final time slice in terms of a deviation function along with the position of the extreme. In this way they set up all 5 degrees of freedom in the Szekeres metric and worked out an algorithm to calculate all other metric functions in the model.

For practical purposes it would be desirable to consider the evolution of inhomogeneities in Szekeres spacetime determined only by initial data. In our approach we specify the curvature function $f$ and the radial density profile with one extreme value at the initial time. Thus we set up only 3 degrees of freedom which is however sufficient to model the density contrast. To make an appropriate choice for the initial density extreme we investigate the shell crossing conditions in terms of the density extremes and show that the resulting constraints are preserved throughout the evolution (for the assumed value of $f$). 

The paper is organized as follows. In the next section we give an overview of the Szekeres metric and describe its basic properties. In section 3 we derive conditions that the density extremes have to satisfy in order to avoid a shell crossing singularity. In section 4 we show that if these conditions are met at the initial time, then they are fulfilled at later times if the curvature function vanishes. In section 5 we derive an analytical formula for the density contrast defined as the difference of the extreme values and for the case $f=0$ we show that if the initial radial inhomogeneity is small, then the density contrast is proportional to ${t^{-3}}$ and the radial dependence of the function does not evolve substantially in time. In section 6 we set up the model by specifying two more functions and show the time evolution of the density contrast in a specific example. We conclude in section 7.

\section{Szekeres spacetime}

The Szekeres spacetime is an exact dust solution of Einstein equations without any symmetries. It was found by Szekeres \cite{2} and generalized by Szafron \cite{3} for an energy momentum tensor describing a perfect fluid.
The LT type of Szekeres metric can be written as \cite{21}
\begin{equation}
\label{9}
{\rm d}s^{2}=-{\rm d}t^{2}+\frac{\left(R'-R\frac{E'}{E}\right)^2}{\epsilon+f\left(r\right)}{\rm d}r^2+\frac{R^2}{E^2}\left({\rm d}p^2+{\rm d}q^2\right),
\end{equation}
where
\begin{equation}
\label{10}
E\left(r,p,q\right)\equiv\frac{S}{2}\left[\left(\frac{p-P}{S}\right)^2+\left(\frac{q-Q}{S}\right)^2+\epsilon\right],
\end{equation}
\begin{equation}\label{25}
E'=\frac{S'}{2}\left[1-\frac{\left(p-P\right)^2}{S^2}-\frac{\left(q-Q\right)^2}{S^2}\right]-\frac{P'}{S}\left(p-P\right)-\frac{Q'}{S}\left(q-Q\right)
\end{equation}
and $f$, $P$, $Q$, $S$ are arbitrary functions of $r$. The parameter $\epsilon$ can have only three  values $-1, 0, 1$ and it determines the geometry of the two-spaces of constant $t$ and $r$.
From Einstein equations it follows a dynamical equation for the function $R$ 
\begin{equation}
\label{11}
{\dot{R}}^2=\frac{2M\left(r\right)}{R}+f\left(r\right)+\frac{\Lambda R^2}{3}
\end{equation}
and an equation for the density evolution
\begin{equation}
\label{12}
\rho=\frac{2}{\kappa c^{2}}\frac{M'-3M\frac{E'}{E}}{R^2\left(R'-R\frac{E'}{E}\right)},
\end{equation}
where prime denotes a derivative with respect to $r$, $M$ is another arbitrary function and $\Lambda$ is cosmological constant. So we have two Einstein equations (\ref{11}), (\ref{12}) and by solving the first one we obtain one more arbitrary function $t_B\left(r\right)$, which enters the solution in the form $t-t_B$. From now on we will use a redefined form of the density $\bar{\rho}\equiv\kappa c^{2}\rho$ and for the sake of simplicity we will drop the bar.

As mentioned above, the metric does not have any symmetries, in other words it has no Killing vectors. Nevertheless the three spaces of constant $t$ are conformally flat \cite{22}.

The parametrization of the metric involves 6 arbitrary functions however the number of physical degrees of freedom is 5, because we can still rescale the radial coordinate (the metric is invariant with respect to the transformation $r'=g(r)$). 

\subsection{Interpretation of $f$ and $t_B$}

The sign of the function $f$ effects the solution of the Einstein equation (\ref{11}). It is a dynamical equation for $R$ and it looks very similar to the Friedman equation, except that here the functions $f$ and $M$ depend on $r$. We will assume only the case when $\Lambda=0$. There are three different types of evolution \cite{23}. If $f<0$ the evolution is elliptic which means that the universe is first expanding and at some point the expansion stops and the universe is collapsing to a final singularity. The solution is given in a parametric form
\begin{equation}
\label{13}
R=\frac{M}{\left(-f\right)}\left(1-\cos\eta\right), \quad \eta-\sin\eta=\frac{\left(-f\right)^{\frac{3}{2}}\left(t-t_B\right)}{M}.
\end{equation}
For $f>0$ the evolution is hyperbolic in the sense that the sign of the expansion does not change, the universe either expands or collapses depending on initial conditions and the parametric solution to the equation (\ref{11}) is
\begin{equation}
\label{14}
R=\frac{M}{\left(-f\right)}\left(\cosh\eta-1\right), \quad \sinh\eta-\eta=\frac{f^{\frac{3}{2}}\left(t-t_B\right)}{M}.
\end{equation}
For $f=0$ the evolution is parabolic and is given by equation
\begin{equation}
\label{15}
R=\left(\frac{9}{2}M\right)^{\frac{1}{3}}\left(t-t_B\right)^{\frac{2}{3}}.
\end{equation}
Since $f$ is a function of $r$, the universe may have different evolution in different regions. The parabolic evolution can be on the boundary between two regions, one having elliptic evolution and the other one hyperbolic. So the sign of $f$ determines the sign of scalar curvature of the three-spaces of constant $t$ and when $f=0=f'$, the three-spaces are flat which corresponds to a presence of pure decaying modes as was shown in \cite{sussman}.

The function $t_B$ is called the bang time function, because $t=t_B$ is the moment when big bang happened. So unlike in homogeneous models, the initial moment of evolution is position dependent. 

The interpretation of the other metric functions ($R$, $M$, $P$, $Q$, $S$ ) depends on $\epsilon$ and from now on we will only consider the case $\epsilon=+1$ which is often called the quasi-spherical case.

\subsection{Coordinate transformation and interpretation of $R$}

We can make a coordinate transformation \cite{11}
\begin{equation}
\label{16}
\frac{p-P}{S}=\cot\frac{\theta}{2}\cos\phi, \quad \frac{q-Q}{S}=\cot\frac{\theta}{2}\sin\phi
\end{equation}
and rewrite the induced two-metric of the surfaces of constant $t$ and $r$
into coordinates $\theta$ and $\phi$. After applying the transformation we get
\begin{equation}
\label{18}
{\rm d}s^2=\frac{4}{S^2}\sin^4\frac{\theta}{2}\left({\rm d}p^2+{\rm d}q^2\right)={\rm d}\theta^2+\sin^2\theta{\rm d}\phi^2,
\end{equation}
which is a metric on a unit sphere. The transformation (\ref{16}) is nothing but a stereographic projection and the function $E$ describes how the $(p,q)$ plane is mapped onto a unit sphere. Every sphere is multiplied by $R^2$ and the function is sometimes called the areal radius, because it is actually the radius of a sphere on a comoving coordinate $r$. 

\subsection{Geometrical meaning of $E$}

In order to understand more about the geometrical properties of the metric (\ref{9}) it is important to investigate the function  $E$. We can see that this function appears in the metric and in the equation for the density (\ref{12}) in the form $\frac{E'}{E}$. Particularly, we can investigate when the function is equal to zero and what are its extreme values. 
Using the transformation (\ref{16}) in (\ref{10}) and (\ref{25}) we can write $\frac{E'}{E}$ as
\begin{equation}
\label{31}
\frac{E'}{E}=-\frac{S'\cos\theta+\sin\theta\left(P'\cos\phi+Q'\sin\phi\right)}{S}.
\end{equation} 
The equation $\frac{E'}{E}=0$ now becomes
\begin{equation}
\label{28}
S'\cos\theta+P'\sin\theta\cos\phi+Q'\sin\theta\sin\phi=0
\end{equation}
and after realizing that 
$\cos\theta=z$, $\sin\theta\cos\phi=y$ and $\sin\theta\sin\phi=x$,
the equation (\ref{28}) becomes
\begin{equation}
\label{30}
S'z+P'x+Q'y=0,
\end{equation}
which is an equation of a plane that goes through the origin of the spherical coordinate system and intersects the $r=t=const$ sphere in a great circle. 
Now calculating the derivatives of (\ref{31}) with respect to $\theta$ and $\phi$ and putting it equal to zero we can find that there are two extreme values and they are 
located at opposite sites on the sphere, in other words if one extreme is at the coordinates $\left(\theta_{1},\phi_{1}\right)$, the other one is at $\left(\pi-\theta_{1}, \phi_{1}+\pi\right)$ and they are located symmetrically with respect to the plane $\frac{E'}{E}=0$, because the coordinates of the extremes are exactly the components of the unit normal to the plane $\frac{E'}{E}=0$. The values of the extremes are
\begin{eqnarray}
\label{33}
\left(\frac{E'}{E}\right)_{max}&=&\frac{\sqrt{\left(S'\right)^2+\left(P'\right)^2+\left(Q'\right)^2}}{S}, \nonumber\\ 
\left(\frac{E'}{E}\right)_{min}&=&-\frac{\sqrt{\left(S'\right)^2+\left(P'\right)^2+\left(Q'\right)^2}}{S},
\end{eqnarray}
where max and min refers to maximum and minimum respectively. So the extremes have opposite values and the function $\frac{E'}{E}$ behaves on the sphere like a dipole \cite{11}. 

\section{Shell crossing conditions in terms of the density extremes}

We are now going to study the formula for the density (\ref{12}). We can see that under certain circumstances the density can change sign or possibly diverge if the denominator becomes zero. Those points where this happens are called the shell crossing singularity. The conditions that the metric functions $M$, $f$, $t_{B}$ and $R$ have to satisfy in order to avoid shell crossing singularity can be found in \cite{11}. In this section we will derive the conditions that the extreme values of the density $\rho_{max}$ and $\rho_{min}$ have to satisfy in order to avoid the shell crossing singularity. 

We can rewrite (\ref{12}) as
\begin{equation}
\label{34}
\rho\left(t,r,\theta, \varphi\right)=2\frac{M'-3M\frac{E'}{E}}{R^{2}\left(R'-R\frac{E'}{E}\right)}=\frac{R'\rho_{LT}-R\frac{E'}{E}\rho_{AV}}{R'-R\frac{E'}{E}},
\end{equation}
where we define
\begin{equation}
\label{35}
\rho_{LT}\left(t,r\right)\equiv\frac{2M'}{R^{2}R'}
\end{equation}
and
\begin{equation}
\label{36}
\rho_{AV}\left(t,r\right)\equiv\frac{6M}{R^{3}}.
\end{equation}
Here $\rho_{LT}$ is the density that we get from (\ref{12}) if we set $\frac{E'}{E}=0$. It is also a radial density in the sense that on a given $r$ it is the value of the density around the great circle that lies in the plane that defines the dipole on the sphere. The index $LT$ refers to Lemaitre-Tolman-Bondi metric, because its density is given with exactly the same formula as (\ref{35}).  
If we make the choice 
\begin{equation}
\label{ic}
R\left(t_{i}\right)\equiv r
\end{equation} 
and use the fact that the function $M$ depends only on $r$, we can express it on the initial time slice when $\frac{E'}{E}=0$ as
\begin{equation}
\label{37}
M=\frac{1}{2}\int_{0}^{r}\rho_{LT0}r'^{2}{\rm d}r',
\end{equation}
where the index $0$ refers to the value of the function at the initial time $t_{i}$.
In that case the definitions $(\ref{35})$ and $(\ref{36})$ can be further rewritten as
\begin{equation}
\label{den}
\rho_{LT}\left(t,r\right) = \frac{1}{R^{2}R'}\rho_{LT0}r^{2}, \quad \quad \rho_{AV}\left(t,r\right) =  \frac{3}{R^{3}}\int_{0}^{r}\rho_{LT0}r'^{2}{\rm d}r'.
\end{equation}
The initial condition (\ref{ic}) and formulas (\ref{37}) - (\ref{den}) will however not be used in the following calculations (with the exception of equation (\ref{68}) giving critical point position) and the main results of this section are independent of it, thus giving us generally applicable criteria. 

%In the definitions $(\ref{35})$ and $(\ref{36})$ we used that the function $M$ depends only on $r$ so we can express it from (\ref{12}) on the initial time slice for the case when $\frac{E'}{E}=0$ as
%\begin{equation}
%\label{37}
%M=\frac{1}{2}\int_{0}^{r}\rho_{LT0}r'^{2}{\rm d}r'
%\end{equation}
%and we made the choice $R\left(t_{i}\right)\equiv r$.
It is good to take a look at the derivative of the density with respect to $\frac{E'}{E}$
\begin{equation}
\label{38}
\frac{\partial\rho}{\partial\frac{E'}{E}}=RR'\frac{\rho_{LT}-\rho_{AV}}{\left(R'-R\frac{E'}{E}\right)^{2}}.
\end{equation}
From the derivative we can see, that if $R'\left(\rho_{LT}-\rho_{AV}\right)>0$ the derivative is positive and the density is growing as $\frac{E'}{E}$ increases. On the other hand if $R'\left(\rho_{LT}-\rho_{AV}\right)<0$ the density will decrease as $\frac{E'}{E}$ increases. Either way we can see that the density behaves on each sphere also like a dipole in the sense that it has two extreme values, maximum and minimum and they are located at the same position as extreme values of the function $\frac{E'}{E}$. In the case when $R'\left(\rho_{LT}-\rho_{AV}\right)<0$ we can write for the density maximum
\begin{equation}
\label{39}
\rho_{max}=\frac{R'\rho_{LT}-R\left(\frac{E'}{E}\right)_{min}\rho_{AV}}{R'-R\left(\frac{E'}{E}\right)_{min}}
\end{equation}
and for the minimum
\begin{equation}
\label{40}
\rho_{min}=\frac{R'\rho_{LT}-R\left(\frac{E'}{E}\right)_{max}\rho_{AV}}{R'-R\left(\frac{E'}{E}\right)_{max}}.
\end{equation}
In the case when $R'\left(\rho_{LT}-\rho_{AV}\right)>0$ the role of $\left(\frac{E'}{E}\right)_{max}$ interchanges with $\left(\frac{E'}{E}\right)_{min}$ in the last two equations.

We split the investigation of the density into three parts, in the first part we investigate when both the numerator and the denominator are positive, in the second case we check when they are both negative and in the last case we take a look at the special case when they are both zero. 
It will be convenient to split the first part into three more subcases depending on the value of $R'\left(\rho_{LT}-\rho_{AV}\right)$. For the investigation we will also need a formula for the function $\frac{E'}{E}$ given in terms of $\rho_{LT}$ and $\rho_{AV}$ which we can derive from (\ref{34})
\begin{equation}
\label{41}
\frac{E'}{E}=\frac{R'}{R}\frac{\rho-\rho_{LT}}{\rho-\rho_{AV}}.
\end{equation}

\vspace{10px}
\noindent\textsc{First case: numerator and denominator of the density are both positive.} \\
\noindent\textsc{Subcase A: $R'\left(\rho_{LT}-\rho_{AV}\right)<0$}

From the positivity of the denominator in (\ref{34}) we get 
\begin{equation}
\label{42}
\frac{R'}{R}>\left(\frac{E'}{E}\right)_{max}=\frac{R'}{R}\frac{\rho_{min}-\rho_{LT}}{\rho_{min}-\rho_{AV}},
\end{equation}
because as follows from (\ref{38}) if $R'\left(\rho_{LT}-\rho_{AV}\right)<0$ we can see that $\frac{E'}{E}$ has maximum where the density has minimum. 
From the dipole property
\begin{equation}
\label{44}
\left(\frac{E'}{E}\right)_{max}=-\left(\frac{E'}{E}\right)_{min}
\end{equation}
we know that $\left(\frac{E'}{E}\right)_{max}$ is non-negative and (\ref{42}) implies that $R'>0$ and therefore we have to consider $\rho_{LT}<\rho_{AV}$. We can now set conditions for $\rho_{min}$ so that the right hand side in (\ref{42}) is nonnegative. 
Inequality (\ref{42}) simplifies to
\begin{equation}
\label{43}
1>\frac{\rho_{min}-\rho_{LT}}{\rho_{min}-\rho_{AV}}
\end{equation}
It is reasonable to require $\rho_{min}\leq\rho_{LT}$, because the minimum should be the smallest value on each sphere ($\rho_{min}=\rho_{LT}$ corresponds to the case when the density is homogeneously distributed on the whole sphere). Since we consider $\rho_{LT}<\rho_{AV}$ it follows that $\rho_{min}<\rho_{AV}$ and the right hand side of (\ref{42}) is nonnegative. It means that besides the obvious condition $\rho_{min}\leq\rho_{LT}$ we don't have any other restriction for the minimum.

Due to the dipole property we can try to derive a condition for the maximum
\begin{equation}
\label{45}
\frac{R'}{R}>\left(\frac{E'}{E}\right)_{max}=-\frac{R'}{R}\frac{\rho_{max}-\rho_{LT}}{\rho_{max}-\rho_{AV}}.
\end{equation} 
Since $\rho_{max}$ should be greater than or equal to $\rho_{LT}$ it implies that $\rho_{max}<\rho_{AV}$ in order for the right hand side in (\ref{45}) to be non-negative. This condition is reasonable since $\rho_{AV}$ is not true value of the density as follows from its definition (\ref{36}). Taking this into consideration we can solve the inequality (\ref{45}) and the solution is
\begin{equation}
\label{46}
\rho_{max}<\frac{1}{2}\left(\rho_{LT}+\rho_{AV}\right)
\end{equation}
and this condition is clearly more restrictive than $\rho_{max}<\rho_{AV}$.

For the numerator of (\ref{34}) we need
\begin{equation}
\label{47}
\frac{R'}{R}\rho_{LT}>\frac{E'}{E}\rho_{AV}
\end{equation}
which can be rewritten as
\begin{equation}
\label{48}
\frac{R'}{R}\frac{\rho_{LT}}{\rho_{AV}}>\left(\frac{E'}{E}\right)_{max}=\frac{R'}{R}\frac{\rho_{min}-\rho_{LT}}{\rho_{min}-\rho_{AV}}.
\end{equation}
It again follows that $R'>0$ and $\rho_{LT}<\rho_{AV}$. The right-hand side of (\ref{48}) has to be non-negative and we know that $\rho_{min}\leq\rho_{LT}$ and since $\rho_{LT}<\rho_{AV}$ we also have $\rho_{min}<\rho_{AV}$ so the inequality (\ref{48}) is always true and we don't get any new condition for $\rho_{min}$. We now use the property (\ref{44}) and look for a condition for $\rho_{max}$
\begin{equation}
\label{49}
\frac{\rho_{LT}}{\rho_{AV}}>-\frac{\rho_{max}-\rho_{LT}}{\rho_{max}-\rho_{AV}}.
\end{equation}
We know that $\rho_{max}\geq\rho_{LT}$ so to make the right hand side non-negative we need $\rho_{max}<\rho_{AV}$. The solution to the inequality (\ref{49}) is
\begin{equation}
\label{50}
\rho_{max}<2\frac{\rho_{AV}\rho_{LT}}{\rho_{LT}+\rho_{AV}}.
\end{equation}
The conditions for the denominator and the numerator have to be satisfied simultaneously, so for the maximum we have (\ref{46}) and (\ref{50}) and (as can be seen from figure 1) the condition (\ref{50}) is stronger in the case of $\rho_{LT}<\rho_{AV}$ and has to be fulfilled in order for the density to be positive. As far as the minimum is concerned, apart from the obvious condition $\rho_{min}\leq\rho_{LT}$ it is not constrained at all.
 
\vspace{10px}
\noindent\textsc{First case: numerator and denominator of the density are both positive.} \\
\noindent\textsc{Subcase B: $R'\left(\rho_{LT}-\rho_{AV}\right)>0$}

The subcase B will be investigated similarly. The only difference is that from (\ref{38}) we know that if $R'\left(\rho_{LT}-\rho_{AV}\right)<0$ the function $\frac{E'}{E}$ has maximum where the density has maximum and this modifies (\ref{39}) and (\ref{40}) accordingly. From the denominator of (\ref{34}) we have
\begin{equation}
\label{51}
\frac{R'}{R}>\left(\frac{E'}{E}\right)_{max}=\frac{R'}{R}\frac{\rho_{max}-\rho_{LT}}{\rho_{max}-\rho_{AV}},
\end{equation} 
which again implies $R'>0$ and $\rho_{LT}>\rho_{AV}$ because $\left(\frac{E'}{E}\right)_{max}\geq 0$ and it simplifies to
\begin{equation}
\label{52}
1>\frac{\rho_{max}-\rho_{LT}}{\rho_{max}-\rho_{AV}}.
\end{equation}
The analysis is basically the same, we again need the right-hand side to be non-negative. We know that $\rho_{max}\geq\rho_{LT}$ and therefore $\rho_{max}>\rho_{AV}$ as well, so the inequality (\ref{52}) will be always true as long as $\rho_{LT}>\rho_{AV}$ and we don't get any further condition for the maximum. For the minimum we have
\begin{equation}
\label{53}
\frac{R'}{R}>\left(\frac{E'}{E}\right)_{max}=-\frac{R'}{R}\frac{\rho_{min}-\rho_{LT}}{\rho_{min}-\rho_{AV}}.
\end{equation} 
In order for the right-hand side to be non-negative we need $\rho_{min}>\rho_{AV}$ which is a valid condition for the density minimum, because as mentioned above $\rho_{AV}$ is not true value of the density. The inequality (\ref{53}) has the solution
\begin{equation}
\label{54}
\rho_{min}>\frac{1}{2}\left(\rho_{AV}+\rho_{LT}\right).
\end{equation}

From the numerator of (\ref{34}) we have the condition
\begin{equation}
\label{55}
\frac{R'}{R}\frac{\rho_{LT}}{\rho_{AV}}>\left(\frac{E'}{E}\right)_{max}=\frac{R'}{R}\frac{\rho_{max}-\rho_{LT}}{\rho_{max}-\rho_{AV}}.
\end{equation}
It again implies $R'>0$ and $\rho_{LT}>\rho_{AV}$ and it does not give us any constraint for the maximum, because we know that $\rho_{max}\geq\rho_{LT}$ and therefore $\rho_{max}>\rho_{AV}$. So (\ref{55}) holds as long as $R'>0$ and $\rho_{LT}>\rho_{AV}$.
For the minimum we can write
\begin{equation}
\label{56}
\frac{R'}{R}\frac{\rho_{LT}}{\rho_{AV}}>\left(\frac{E'}{E}\right)_{max}=-\frac{R'}{R}\frac{\rho_{min}-\rho_{LT}}{\rho_{min}-\rho_{AV}}.
\end{equation}
The right-hand side will be positive if $\rho_{min}>\rho_{AV}$ and the solution to (\ref{56}) is
\begin{equation}
\label{57}
\rho_{min}>2\frac{\rho_{LT}\rho_{AV}}{\rho_{LT}+\rho_{AV}}.
\end{equation}
All together in the case of $R'\left(\rho_{LT}-\rho_{AV}\right)>0$ we have two conditions (\ref{54}) and (\ref{57}) for the minimum and as can be seen from figure 1 the first one is stronger. For the maximum, except for the obvious $\rho_{max}\geq\rho_{LT}$, we don't have any constraint so the density will be positive as long as (\ref{54}) holds and $\rho_{max}\geq\rho_{LT}$.

\begin{figure}[!h]
\centering
\includegraphics[scale=0.5]{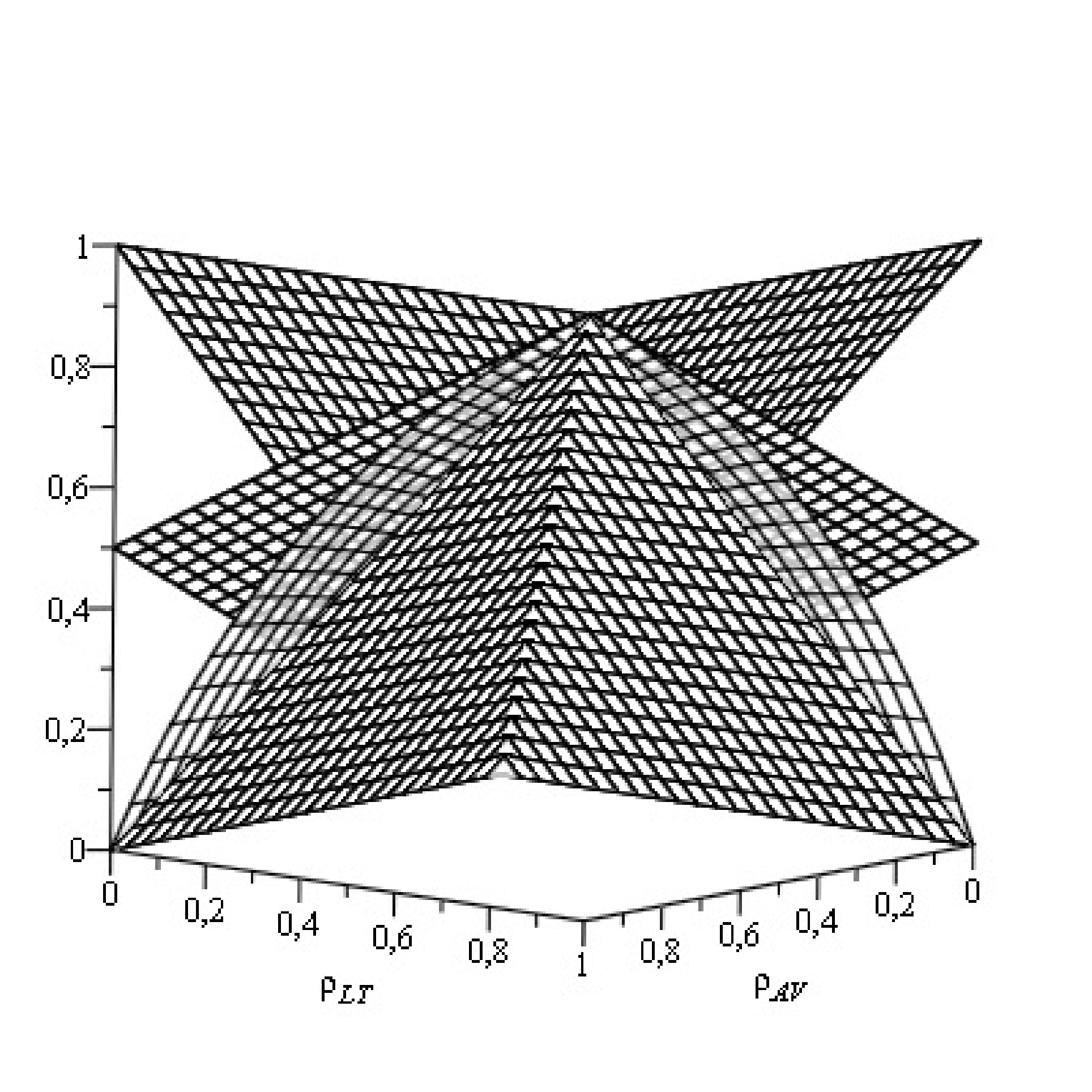}
\caption{In this figure we can see 4 different surfaces that constrain the density extremes. In the right part of the figure they are from the bottom to the top: $\rho_{AV}$,  $2\frac{\rho_{AV}\rho_{LT}}{\rho_{AV}+\rho_{LT}}$, $\frac{1}{2}\left(\rho_{AV}+\rho_{LT}\right)$, $\rho_{LT}$.}
\end{figure}

\vspace{10px}
\noindent\textsc{First case: numerator and denominator of the density are both positive.}\\
\noindent\textsc{Subcase C: $R'\left(\rho_{LT}-\rho_{AV}\right)=0$}

Before we consider this special case we can first rewrite (\ref{34}) as
\begin{equation}
\rho=\frac{R'\left(\rho_{LT}-\rho_{AV}\right)+\rho_{AV}\left(R'-R\frac{E'}{E}\right)}{R'-R\frac{E'}{E}}
\end{equation}
Now using $R'\left(\rho_{LT}-\rho_{AV}\right)=0$ we can see that the formula for the density simplifies
\begin{equation}
\rho=\rho_{AV}\frac{R'-R\frac{E'}{E}}{R'-R\frac{E'}{E}}=\rho_{AV},
\end{equation}
in other words the value of the density is independent of $\theta$ and $\phi$ and the density is homogeneously distributed on the whole sphere as follows also from (\ref{38}) because the derivative of the density is now zero. Using the same argument as in subcase A and B we have $R'>0$ and therefore $\rho_{LT}=\rho_{AV}$. For the extreme values of the density we obviously need in this case
\begin{equation}
\rho_{max}=\rho_{min}=\rho_{AV}=\rho_{LT}.
\end{equation}

%In the case when $R'=0$ and $\frac{E'}{E}\neq 0$ we can see from (\ref{34}) that the formula for the density also simplifies 
%\begin{equation}
%\rho=\frac{R\frac{E'}{E}\rho_{AV}}{R\frac{E'}{E}}=\rho_{AV}
%\end{equation}
%and the density is again homogeneous on the whole sphere and obviously it follows that $\rho_{LT}=\rho_{AV}=\rho_{max}=\rho_{min}$. The case when both $R'=0$ and $\frac{E'}{E}=0$ will be considered separately in the third case.

The point where $\rho_{LT}=\rho_{AV}$ is especially interesting because the shell-crossing conditions for $\rho_{max}$ and $\rho_{min}$ change here and the location of $\rho_{max}$ and $\rho_{min}$ interchanges by passing through the plane $\frac{E'}{E}=0$. The position of this critical point $r_{c}$ generally evolves and we can get its value as the solution to the equation
\begin{equation}
\label{68}
3\frac{R'}{R}\int_{0}^{r_{c}}\rho_{LT0}r^{2}{\rm d}r=\rho_{LT0}r_{c}^{2},
\end{equation}
where we used the initial condition (\ref{ic}) and formulas (\ref{den}) in order to express $\rho_{LT}$ and $\rho_{AV}$. So the time evolution of the critical point will depend on the time evolution of the function $\frac{R'}{R}$.

\vspace{10px}
\noindent\textsc{Second case: numerator and denominator of the density are both negative.}

In order for the denominator of the density (\ref{34}) to be negative, we need
\begin{equation}
\label{58}
\frac{R'}{R}<\left(\frac{E'}{E}\right)_{min}.
\end{equation}
This condition will not be fulfilled unless $R'<0$ because $\left(\frac{E'}{E}\right)_{min}\leq 0$. The investigation is than similar as in the first case and leads to the same conditions for $\rho_{min}$ and $\rho_{max}$ as can be easily verified. 

On the other hand if $R'>0$ and we allow the denominator to be negative on a given $r$, it will be also positive on that $r$ for some specific $\theta$ and $\phi$. This behaviour may or may not be correct, depending on the behaviour of the numerator and specifically depending on whether or not the numerator changes sign at the same $\theta$, $\phi$. For the numerator we have inequality
\begin{equation}
\label{59}
R'\rho_{LT}-R\frac{E'}{E}\rho_{AV}<0.
\end{equation}
This can be rewritten as
\begin{equation}
\label{60}
\frac{R'}{R}\frac{\rho_{LT}}{\rho_{AV}}<\left(\frac{E'}{E}\right)_{min}.
\end{equation}
The same argument as in the case of denominator tells us that this will be satisfied only if $R'<0$. So if the numerator is negative on a given $r$ but $R'>0$, there is a region on this $r$ where it is positive too.
% The question now is, if the numerator and denominator change the sign at the same $\theta$, $\phi$. 
For the positiveness of the denominator in a region $(\theta, \phi)$ we have
\begin{equation}
\label{61}
\frac{R'}{R}>\frac{E'}{E}
\end{equation}
and for the numerator the condition is
\begin{equation}
\label{62}
\frac{R'}{R}\frac{\rho_{AV}}{\rho_{LT}}>\frac{E'}{E}.
\end{equation}
Clearly the conditions are not the same unless $\rho_{LT}=\rho_{AV}$, so except for this special case, there will always be a region on a given $r$ where the numerator and the denominator will have different signs. The places where the sign changes are in both cases circles on the sphere, that are parallel to the great circle that defines the plane of the dipole. The order in which the circles go are seen in figure 2, where the parallel circles are mapped as parallel lines. 

So if in the second case $R'>0$ there are no conditions for the density extremes that would prevent shell crossing.

\begin{figure}[!h]
\centering
\subfloat[]{\includegraphics[trim = 10mm 170mm 110mm 20mm, clip, width=6cm]{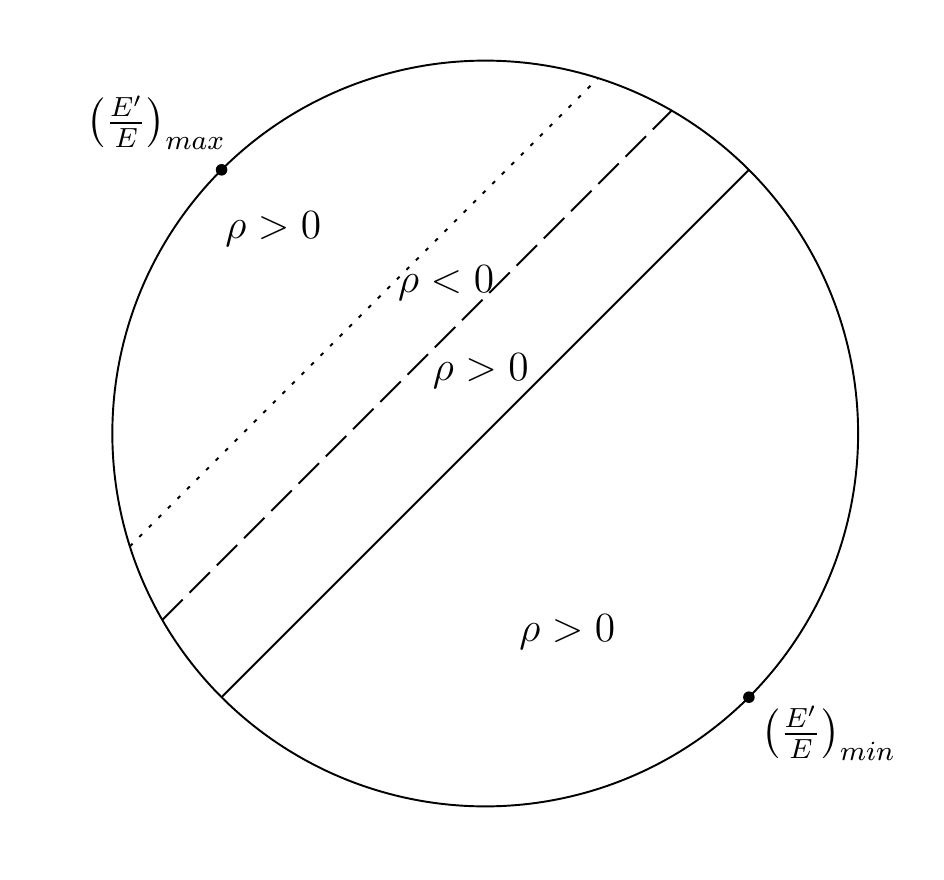}}
%\newline
\subfloat[]{\includegraphics[trim = 10mm 170mm 110mm 20mm, clip, width=6cm]{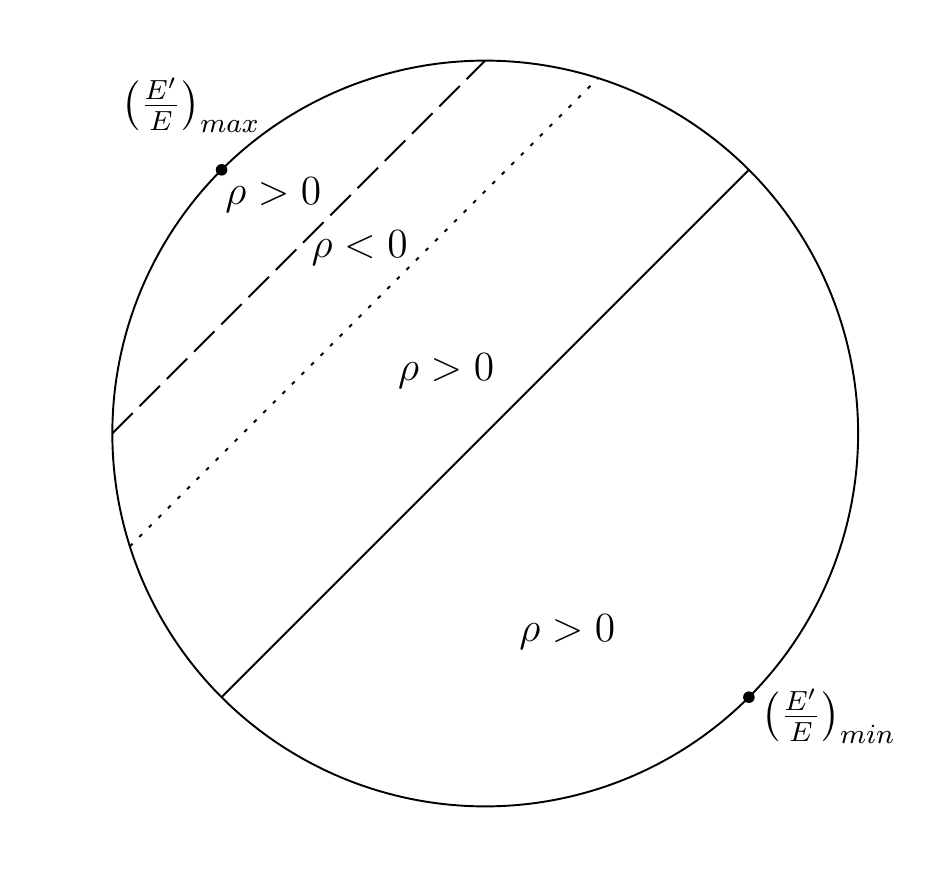}}
\caption{The behaviour of the density for $\rho_{AV}>\rho_{LT}$ in panel (a) and $\rho_{AV}<\rho_{LT}$ in panel (b) when the conditions for no shell-crossing are not met. The circle represents a sphere on a given $r$, the solid line represents the plane $\frac{E'}{E}=0$. The dashed line represents a plane $\frac{E'}{E}\frac{\rho_{LT}}{\rho_{AV}}=\frac{R'}{R}$, that is parallel to the plane $\frac{E'}{E}=0$, the density becomes zero here. The dotted line is the plane $\frac{E'}{E}=\frac{R'}{R}$ and it is the plane where the density diverges. The density is negative between the dashed and dotted line.}
\label{fig_2}
\end{figure}

\vspace{10px}
\noindent\textsc{Third case: numerator and denominator of the density are both zero.}

The denominator is zero when
\begin{equation}
\label{63}
\frac{R'}{R}=\frac{E'}{E},
\end{equation}
but this equation will be satisfied on a given $r$ for all $\theta$ and $\phi$ only if both $\frac{R'}{R}$ and $\frac{E'}{E}$ are zero. This consequently means that $E'=P'=Q'=S'=R'=0$. The same arguments are valid for the numerator of the density. As was discussed in \cite{11} and in the context of LTB metric in \cite{23} this point will not be a singularity if also $M'=0$. Since $M'$ and $E'$ are independent of time it implies that $R'$ cannot evolve and we need stationary configuration in order to avoid shell crossing.  

\begin{table}[!ht]
\label{tab1}
\centering
\begin{tabular}{|c| c|}
\hline
Conditions for $\rho_{min}$  & Conditions for $\rho_{max}$ \\
\hline
\multicolumn{2}{|c|}{$\rho_{AV}>\rho_{LT}$}      \\
\hline
$\rho_{min}\leq\rho_{LT}<\rho_{AV}$ & $\rho_{max} \in\langle\left.\rho_{LT}; 2\frac{\rho_{AV}\rho_{LT}}{\rho_{LT}+\rho_{AV}}\right) $\\

\hline
\multicolumn{2}{|c|}{$\rho_{AV}<\rho_{LT}$}      \\
\hline
%\centering
$\rho_{min}\in\left(\frac{\rho_{AV}+\rho_{LT}}{2}; \rho_{LT}\right.\rangle$ & $\rho_{max}\geq\rho_{LT}>\rho_{AV}$\\
\hline
\multicolumn{2}{|c|}{$\rho_{AV}=\rho_{LT}$}      \\
\hline
%\centering
$\rho_{min}=\rho_{LT}$ & $\rho_{max}=\rho_{LT}$\\
\hline
\end{tabular}
\caption{The list of shell crossing conditions for density extremes in terms of $\rho_{LT}$ and $\rho_{AV}$.}
\end{table}

The table 1 summarizes the conditions for the extreme values of the density that we obtained so that shell crossing would be avoided. From the dipole property (\ref{44}) it follows that there is a constraint between the density extremes
\begin{equation}
\label{69}
\frac{\rho_{min}-\rho_{LT}}{\rho_{min}-\rho_{AV}}=-\frac{\rho_{max}-\rho_{LT}}{\rho_{max}-\rho_{AV}},
\end{equation}
which means that knowing one extreme value allows us calculate the other one with a formula
\begin{equation}
\label{70}
\rho_{max}=\frac{\rho_{min}\left(\rho_{AV}+\rho_{LT}\right)-2\rho_{LT}\rho_{AV}}{2\rho_{min}-\rho_{LT}-\rho_{AV}}
\end{equation}
that follows from (\ref{69}). And for the other extreme we just interchange min for max in the last equation. Consequently it is sufficient to ensure that just one of the extremes satisfies the conditions in table 1 and the constraint (\ref{69}) ensures that the other density extreme has a value that does not break the conditions in table 1.

%\begin{figure}[!ht]
%\includegraphics[trim = 10mm 170mm 110mm 20mm, clip, width=8cm]{obr2.pdf}
%\end{figure}
%\begin{figure}[!ht]
%\includegraphics[trim = 10mm 170mm 110mm 20mm, clip, width=8cm]{obr.pdf}
%\caption{In this figure we can see 4 different surfaces that constrain the density extremes. In the right part of the figure they are from the bottom to the top: $\rho_{AV}$,  $2\frac{\rho_{AV}\rho_{LT}}{\rho_{AV}+\rho_{LT}}$, $\frac{1}{2}\left(\rho_{AV}+\rho_{LT}\right)$, $\rho_{LT}$.}
%\end{figure}

\section{Time dependence of the shell crossing conditions, case $f=0$}

In the previous section we derived conditions that $\rho_{max}$ and $\rho_{min}$ have to satisfy in order to avoid shell crossing singularity. To find out if a shell crossing occurs at a given time, we need to calculate $\rho_{max}$ or $\rho_{min}$ at that time and then check if the conditions are met. It would be useful to have conditions in terms of the initial $\rho_{LT0}$ and $\rho_{min0}$ or $\rho_{max0}$ that would ensure no shell crossing at any time during the evolution. In this section we show that in the special case when $f=0$, if we avoid shell crossing at the initial time, it is guaranteed that no shell crossing occurs during the time evolution. 

First we show that if on the initial time slice $\rho_{AV0}>\rho_{LT0}$, then this condition holds at any later time. We will use the initial condition (\ref{ic}) and start with the inequality that we want to prove $\rho_{AV}>\rho_{LT}$
\begin{equation}
\label{71}
\frac{3}{R^{3}}\int_{0}^{r}\rho_{LT0}r'^{2}{\rm d}r'>\frac{1}{R^{2}R'}\rho_{LT0}r^2,
\end{equation}
this can be rewritten as
\begin{equation}
\label{72}
\frac{R'}{R}r\,\frac{3\int_{0}^{r}\rho_{LT0}r'^{2}{\rm d}r'}{\rho_{LT0}r^{3}}>1
\end{equation}
and the inequality sign depends on the sign of $R'$ and here we assumed $R'>0$. We can see that the time dependence in the last inequality is hidden in the function $\frac{R'}{R}r$. 
The choice $f=0$ has parabolic evolution and the solution to the equation (\ref{11}) is given by (\ref{15}). If we substitute in (\ref{15}) for $M$ from (\ref{37}) we get
\begin{equation}
\label{73}
R\left(t,r\right)=\left(\frac{9}{4}\right)^{\frac{1}{3}}\left(\int_{0}^{r}\rho_{LT0}r'^{2}{\rm d}r'\right)^{\frac{1}{3}}\left(t-t_{B}\right)^{\frac{2}{3}}.
\end{equation}
We can fix the bang time function $t_{B}$ using (\ref{ic}) to be
\begin{equation}
\label{74}
t_{B}=\left[t_{i}-\frac{2}{3}\left(\frac{r^{3}}{\int_{0}^{r}\rho_{LT0}r'^{2}{\rm d}r'}\right)^{\frac{1}{2}}\right],
\end{equation}
where $t_{i}$ is the initial moment of the evolution. By calculating the radial derivative of (\ref{73}) we can express $\frac{R'}{R}r$ as
\begin{equation}
\label{75}
\frac{R'}{R}r=\frac{\rho_{LT0}r^{3}}{3\int_{0}^{r}\rho_{LT0}r'^{2}{\rm d}r'}-\frac{2}{3}\frac{t_{B}'}{t-t_{B}}r.
\end{equation}
After calculating radial derivative of the bang time function (\ref{74}) and substituting it in the last formula we obtain
\begin{equation}\label{76}
\frac{R'}{R}r=\frac{\rho_{LT0}r^{3}}{3\int_{0}^{r}\rho_{LT0}r'^{2}{\rm d}r'}+\frac{2}{3}\frac{\frac{3\int_{0}^{r}\rho_{LT0}r'^{2}{\rm d}r'}{\rho_{LT0}r^{3}}-1}{\left(\frac{3\int_{0}^{r}\rho_{LT0}r'^{2}{\rm d}r'}{\rho_{LT0}r^{3}}\right)^{\frac{3}{2}}}\,\frac{1}{t-t_{B}}\sqrt{\frac{3}{\rho_{LT0}}}.
\end{equation}
Now we define
\begin{equation}
\label{77}
A\equiv\frac{\rho_{AV0}}{\rho_{LT0}}=\frac{3\int_{0}^{r}\rho_{LT0}r'^{2}{\rm d}r'}{\rho_{LT0}r^{3}}
\end{equation}
and using this definition we can rewrite (\ref{76}) as
\begin{equation}
\label{78}
\frac{R'}{R}r=\frac{1}{A}+\frac{2}{3}\frac{A-1}{A^{\frac{3}{2}}}\frac{1}{t-t_{B}}\sqrt{\frac{3}{\rho_{LT0}}}.
\end{equation}
From this formula it can be shown that $\frac{R'}{R}r$ is a monotonic function of $t$. For $t=t_{i}$, its value is 1, which follows from our initial condition $R\left(t_{i}\right)=r$ and for $t\to\infty$, $\frac{R'}{R}r\to\frac{1}{A}$ so it is increasing if $\rho_{AV0}>\rho_{LT0}$ and it is decreasing if $\rho_{AV0}<\rho_{LT0}$. In any case the function is positive which implies $R'>0$ so our assumption of the inequality sign in (\ref{72}) is justified.
Using the definition (\ref{77}) and substituting for $\frac{R'}{R}r$ into (\ref{72}) we get
\begin{equation}
\label{79}
\frac{2}{3}\frac{A-1}{\sqrt{A}}\frac{1}{t-t_{B}}\sqrt{\frac{3}{\rho_{LT0}}}>0.
\end{equation} 
Since $t>t_{B}$ the last inequality will be satisfied as long as $A>1$, which is equivalent to $\rho_{AV0}>\rho_{LT0}$ as can be seen from the definition of $A$. So we can see exactly what we wanted to prove, if $\rho_{AV0}>\rho_{LT0}$ than $\rho_{AV}>\rho_{LT}$ at any time. Similarly we could prove that if $\rho_{AV0}<\rho_{LT0}$ than  $\rho_{AV}<\rho_{LT}$ at any time.  Next we will assume that $\rho_{AV0}>\rho_{LT0}$ and we will express the shell crossing condition $\rho_{min}<\rho_{LT}$ in terms of initial data.
For the function $\frac{E'}{E}$ we have formula (\ref{41}) and since the function does not depend on time, we can express it on the initial time slice
\begin{equation}
\label{80}
\frac{E'}{E}_{max}=\frac{1}{r}\frac{\rho_{min0}-\rho_{LT0}}{\rho_{min0}-\rho_{AV0}}.
\end{equation}
We plug the last formula into (\ref{40}) and after some calculations we get
\begin{equation}
\label{81}
\rho_{min}=\frac{r^{3}}{R^{3}}\rho_{LT0}\frac{\frac{\rho_{min0}}{\rho_{LT0}}\left(1-\frac{\rho_{AV0}}{\rho_{LT0}}\right)}{\frac{R'}{R}r\left(\frac{\rho_{min0}}{\rho_{LT0}}-\frac{\rho_{AV0}}{\rho_{LT0}}\right)-\frac{\rho_{min0}}{\rho_{LT0}}+1}
\end{equation}
and we used $\frac{\rho_{AV}}{\rho_{LT}}=\frac{R'}{R}r\frac{\rho_{AV0}}{\rho_{LT0}}$ which follows from its definitions (\ref{35}) and (\ref{36}).
We will now define
\begin{equation}
\label{82}
C\equiv\frac{\rho_{min0}}{\rho_{LT0}}
\end{equation}
and using this definition with (\ref{77}) we can rewrite (\ref{81}) as 
\begin{equation}
\label{83}
\rho_{min}=\frac{r^{3}}{R^{3}}\rho_{LT0}\frac{C\left(1-A\right)}{\frac{R'}{R}r\left(C-A\right)+\left(1-C\right)}.
\end{equation}
%When we substitute for $R$ from (\ref{73}) in the last formula we get
%\begin{equation}
%\label{84}
%\rho_{min}=\frac{4}{3}\frac{1}{\left(t-t_{B}\right)^{2}}\frac{1}{A}\frac{C\left(1-A\right)}{\frac{R'}{R}r\left(C-A\right)+\left(1-C\right)}.
%\end{equation}
The inequality $\rho_{min}<\rho_{LT}$ can now be rewritten as
%\begin{equation}
%\label{85}
%\frac{4}{3}\frac{1}{\left(t-t_{B}\right)^{2}}\frac{1-A}{A}\frac{C}{\frac{R'}{R}r\left(C-A\right)+\left(1-C\right)}<\frac{1}{R'R^{2}}\rho_{LT0}r^{2},
%\end{equation}
%which after substituting for $R$ and some more calculations simplifies to
\begin{equation}
\label{86}
\frac{R'}{R}r\frac{\left(1-A\right)C}{\frac{R'}{R}r\left(C-A\right)+\left(1-C\right)}<1.
\end{equation}
We need to multiply the last inequality by the denominator, but in order to do that we need to find out its sign. 
From (\ref{77}) and (\ref{82}) it follows
\begin{equation}
C-A=-\frac{1}{\rho_{LT0}}\left(\rho_{AV0}-\rho_{min0}\right)<0,
\end{equation}
next using the minimal value of $\frac{R'}{R}r$ to be $\frac{1}{A}$ we can write for the denominator of (\ref{86})
\begin{equation}\label{87}
\left(1-C\right)-\frac{R'}{R}r\left(A-C\right)\leq 1-C-\frac{1}{A}\left(A-C\right)=C\left(\frac{1}{A}-1\right)<0,
\end{equation}
because we assume $\rho_{AV0}>\rho_{LT0}$, i.e $A>1$. So the denominator is negative and when we multiply by it in (\ref{86}), we get after some calculations
\begin{equation}
\label{88}
\frac{R'}{R}rA>1,
\end{equation}
%which means
%\begin{equation}
%\label{89}
%\frac{2}{3}\frac{A-1}{\sqrt{A}}\frac{1}{t-t_{B}}\sqrt{\frac{3}{\rho_{LT0}}}>0
%\end{equation}
which is inequality (\ref{72}) written in terms of $A$ and we already proved that it holds as long as $A>1$. So we can see that if $\rho_{AV0}>\rho_{LT0}$ than $\rho_{min}<\rho_{LT}$ is fulfilled and the shell crossing will be avoided at any time. Similarly it could be shown that in the case $\rho_{AV0}<\rho_{LT0}$, i.e. $A<1$, if $\rho_{max0}>\rho_{LT0}$ than $\rho_{max}>\rho_{LT}$ at any time.

We can also see that if $\rho_{AV0}=\rho_{LT0}$ at some point $r_{c}$, it means that we have $A=1$ and from (\ref{78}) it follows $\frac{R'}{R}r=1$. So the position of the critical point $r_{c}$ as defined in (\ref{68}) is time independent.

\section{Density contrast, case $f=0$}

We will choose $\rho_{LT0}$ in such a way that $\rho_{LT0}<\rho_{AV0}$. We already expressed $\rho_{min}$ in terms of $\rho_{LT}$ and $\rho_{min0}$ in (\ref{83}). Similarly we can express $\rho_{max}$ in terms of $\rho_{LT0}$ and $\rho_{max0}$ as
\begin{equation}
\label{90}
\rho_{max}=\frac{r^{3}}{R^{3}}\rho_{LT0}\frac{D\left(1-A\right)}{\frac{R'}{R}r\left(D-A\right)-\left(D-1\right)},
\end{equation}
where we define
\begin{equation}
\label{91}
D\equiv\frac{\rho_{max0}}{\rho_{LT0}}.
\end{equation}

We are now interested in the time evolution of the difference of the extreme values of the density, so we define
\begin{eqnarray}
\label{92}
\Delta\rho\equiv\rho_{max}-\rho_{min}=\frac{r^{3}}{R^{3}}\rho_{LT0}\left(1-A\right)\times\nonumber\\
\left[\frac{D}{\frac{R'}{R}r\left(D-A\right)-\left(D-1\right)}-\frac{C}{\frac{R'}{R}r\left(C-A\right)-\left(C-1\right)}\right]\ 
\end{eqnarray}
We can now substitute for $R$ from (\ref{73}) and after some more calculations it can be rewritten in the form
\begin{equation}
\label{93}
\Delta\rho\left(t,r\right)=\frac{4}{3}\frac{1}{\left(t-t_{B}\right)^{2}}\frac{\Delta\rho_{0}}{\rho_{LT0}}h\left(t,r\right)\, ,
\end{equation}
where we have defined $\Delta\rho_{0}\equiv\rho_{max0}-\rho_{min0}$,
\begin{equation}\label{94}
h\left(t,r\right)\equiv\frac{1-A}{A}\ \frac{1-\frac{R'}{R}rA}{\left(\frac{R'}{R}r\right)^{2}\left(D-A\right)\left(C-A\right)+\left(D-1\right)\left(C-1\right)},
\end{equation}
and we used
\begin{equation}
\label{95}
2\left(CD+A\right)-\left(C+D\right)\left(A+1\right)=0,
\end{equation}
which can be derived from the constraint (\ref{70}). We will now investigate the behaviour of the function $h$. 

We set $A=1+\epsilon$ and take a look at how the function $h$ behaves in the case when $\epsilon<<1$, which corresponds to the situation when the radial derivative of $\rho_{LT0}$ is small. First we approximate $\frac{R'}{R}r$ that is given by (\ref{78}),
\begin{eqnarray}
\label{96}
\frac{R'}{R}r&=&\frac{1}{1+\epsilon}\left(1+\frac{2}{3}\frac{1}{t-t_{B}}\sqrt{\frac{3}{\rho_{LT0}}}\frac{\epsilon}{\sqrt{1+\epsilon}}\right)\approx\\
&\approx& 1+\left(\frac{2}{3}\frac{1}{t-t_{B}}\sqrt{\frac{3}{\rho_{LT0}}}-1\right)\epsilon+\left(2-\frac{1}{3}\frac{1}{t-t_{B}}\sqrt{\frac{3}{\rho_{LT0}}}\right)\epsilon^{2}+o\left(\epsilon^{3}\right).\nonumber
\end{eqnarray}
So in the numerator on the right hand side of (\ref{94}) we get
\begin{equation}\label{97}
\left(1-A\right)(1-\frac{R'}{R}rA)=-\left(1-\frac{R'}{R}r\right)\epsilon+\frac{R'}{R}r\epsilon^{2}\approx\frac{2}{3}\frac{1}{t-t_{B}}\sqrt{\frac{3}{\rho_{LT0}}}\epsilon^{2}+o\left(\epsilon^{3}\right).
\end{equation}
In order to simplify the denominator in (\ref{94}) we set $D=1+\delta$ and $C=1-\xi$ and we assume that $\Delta\rho_{0}<<\rho_{LT0}$ in which case $\delta<<1$ and $\xi<<1$. In this case the approximation of the denominator reads
\begin{eqnarray}
\label{98}
&&A\left[\left(\frac{R'}{R}r\right)^2\left(C-A\right)\left(D-A\right)+\left(C-1\right)\left(D-1\right)\right]\approx\nonumber\\
&&\quad\approx\left(C-A\right)\left(D-A\right)+\left(C-1\right)\left(D-1\right)
\end{eqnarray}
and for the function $h$ we get
\begin{equation}
\label{99}
h\approx \frac{2}{\sqrt{3}}\frac{1}{t-t_{B}}\frac{1}{\sqrt{\rho_{LT0}}}\frac{\left(A-1\right)^{2}}{\left(C-A\right)\left(D-A\right)+\left(C-1\right)\left(D-1\right)}.
\end{equation}
%The expression $\left(C-A\right)\left(D-A\right)+\left(C-1\right)\left(D-1\right)$ that is in (\ref{99}) can be rewritten as
%\begin{eqnarray}
%\label{100}
%&&\left(C-A\right)\left(D-A\right)+\left(C-1\right)\left(D-1\right)=\nonumber\\
%&=&2CD-AC-AD+A^{2}-C-D+1.
%\end{eqnarray}
From (\ref{95}) it follows
\begin{equation}
\label{101}
2CD=AC+AD-2A+C+D.
\end{equation}
Using (\ref{91}) we can rewrite the denominator in (\ref{99})
\begin{equation}\label{102}
\left(C-A\right)\left(D-A\right)+\left(C-1\right)\left(D-1\right)=A^{2}-2A+1=\left(A-1\right)^{2}.
\end{equation}
So for $h$ we can write
\begin{equation}
\label{103}
h\approx\frac{2}{\sqrt{3}}\frac{1}{t-t_{B}}\frac{1}{\sqrt{\rho_{LT0}}}
\end{equation}
and for $\Delta\rho$ we have
\begin{equation}
\label{104}
\Delta\rho\approx\frac{8\sqrt{3}}{9}\frac{1}{\left(t-t_{B}\right)^{3}}\frac{\Delta\rho_{0}}{\rho_{LT0}^{\frac{3}{2}}}.
\end{equation}
From the last equation we can see that if the radial derivative of $\rho_{LT0}^{-\frac{3}{2}}$ is small then it does not effect much the shape of the initial difference of the density extremes. Also if at late time $t$ the bang time function is small compared to $t$ we can see that $\Delta\rho$ is proportional to $\frac{1}{t^{3}}$. We can try to evaluate (\ref{104}) at the initial time $t=t_{i}$
\begin{equation}
\label{105}
\Delta\rho\left(t=t_{i}\right)\approx\frac{8\sqrt{3}}{9}\frac{27}{8}\frac{\left(\int\rho_{LT0}r^{2}{\rm d}r\right)^{\frac{3}{2}}}{r^{\frac{9}{2}}}\frac{\Delta\rho_{0}}{\rho_{LT0}^{\frac{3}{2}}}=A^{\frac{3}{2}}\Delta\rho_{0},
\end{equation}
so we can see that if A is close to 1 the approximation gives us what we expect.

\section{Model specification}

The Szekeres spacetime has 5 degrees of freedom, so to fully specify the model we need to set up 5 functions. But since we are only interested in the density contrast, it is sufficient to specify just three functions, which is the curvature function $f$ that we assume to be zero, the initial radial profile $\rho_{LT0}$ and one extreme value of the density. We will set up two models $a$ and $b$. For the initial radial profile we choose 
\begin{eqnarray}
\label{106}
&a:&\quad\rho_{LT0}=\rho_{b0}\left(1+\frac{1}{10}e^{-\frac{r^{2}}{500}}\right),\nonumber\\
&b:&\quad\rho_{LT0}=\rho_{b0}\left(1+\frac{1}{5}e^{-\frac{r^{2}}{500}}\right).
\end{eqnarray}
Both of them are peaked at the origin and $\rho_{b0}$ is the background density at the initial time. The $\rho_{LT0}$ is chosen in such a way that besides the origin it is less then $\rho_{AV0}$, therefore for the extreme value we specify the density minimum since the only requirement for it is that it has to be equal to $\rho_{LT0}$ at the origin and less everywhere else. We choose $\rho_{min0}$ as
\begin{eqnarray}
\label{107}
&a:&\quad\rho_{min0}=\rho_{b0}\left(1+\frac{1}{10}e^{-\frac{r^{2}}{400}}\right),  \nonumber\\
&b:&\quad \rho_{min0}=\rho_{b0}\left(1+\frac{1}{5}e^{-\frac{r^{2}}{400}}\right).
\end{eqnarray}
The models differ by the size of the radial inhomogeneity that is in the model $b$ twice as big as in the model $a$ and we want to demonstrate, that the approximation formula will work better for the model $a$, because the function $A$ is closer to 1 than in the model $b$. The functions $\rho_{LT0}$, $\rho_{min0}$, $\rho_{max0}$ and $\rho_{AV0}$ for the model $a$ are shown in figure 3a. The function $A$ for both models is shown in figure 3b. The initial time is chosen as $t_{i}=5\cdot 10^{5}y$ that approximately corresponds to the time of last scattering. The final time $t_{f}=13.7\cdot 10^{9}y$ which is approximately present time.  The initial density contrast $\frac{\Delta\rho_{0}}{\rho_{b0}}$ and the final density contrast $\frac{\Delta\rho}{\rho_{b}}$ for both models are shown in figures 4,  $\rho_{b}$ denotes the background density at the final time. The impact of the time evolution on the initial shape is minimal as we expected since $A$ is close to $1$, on the other hand
  the magnitude of the density contrast drops significantly because of the factor $t^{-3}$ in (\ref{104}). We can also see in the figure 4 that the formula (\ref{104}) approximates the density contrast better for the model $a$, that corresponds to the lower peak, because we chose the radial inhomogeneity lower than in model $b$.  

%\begin{figure}[!ht]
%\begin{minipage}{8cm}
%\includegraphics[scale=0.35]{plot1.eps}
%\caption{The functions $\rho_{min0}$, $\rho_{max0}$, $\rho_{LT0}$ and $\rho_{AV0}$ for the model $a$. The dotted line represents the chosen initial density minimum $\rho_{min0}$ according to (\ref{107}). The dashed line is the initial density maximum $\rho_{max0}$ computed according to (\ref{70}). The solid line between the dotted and dashed lines is the chosen initial radial density profile $\rho_{LT0}$ as specified in (\ref{106}). The top solid line is the computed $\rho_{AV0}$. All values on the vertical axis are divided by $\rho_{b}$.}
%\end{minipage}
%\quad
%\begin{minipage}{8cm}
%\vspace{-100pt}
%\includegraphics[scale=0.35]{A.eps}
%\caption{The function $A$ as defined in (\ref{77}) for both models. The lower curve corresponds to the model $a$, the upper curve corresponds to the model $b$.}
%\end{minipage}
%\end{figure}

\begin{figure}[]
\subfloat[]{\includegraphics[scale=0.3]{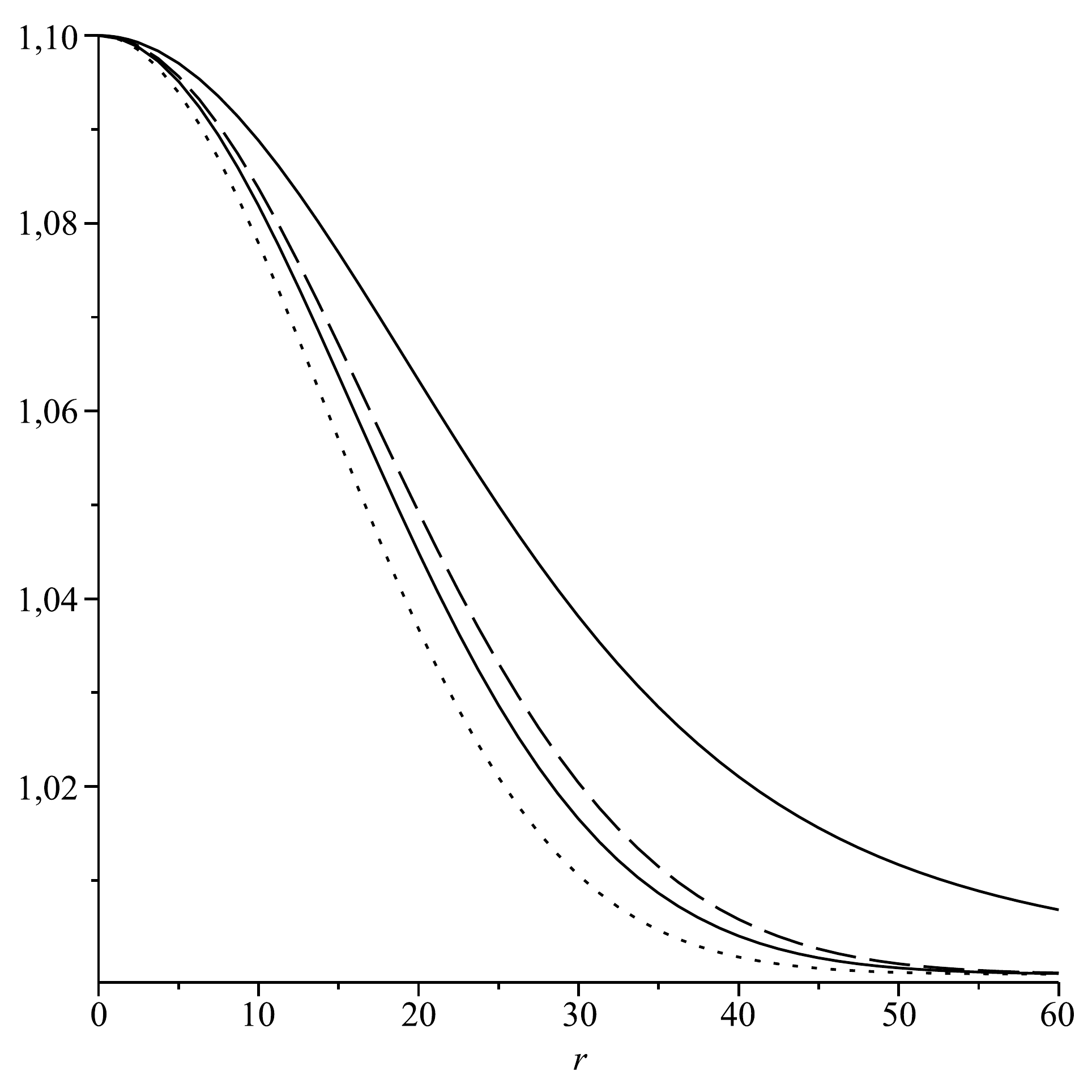}}
\subfloat[]{\includegraphics[scale=0.3]{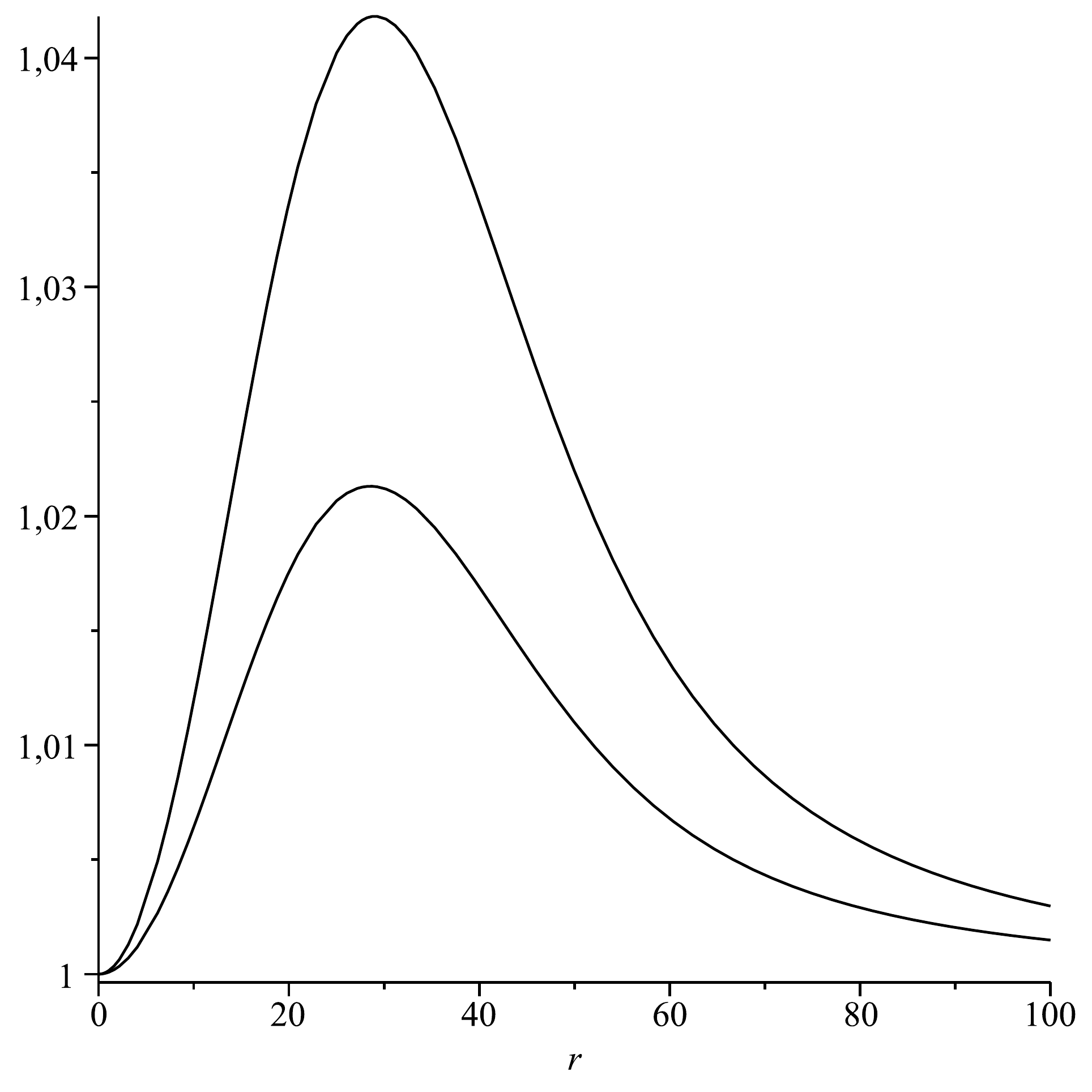}}
\caption{Panel (a): The functions $\rho_{min0}$, $\rho_{max0}$, $\rho_{LT0}$ and $\rho_{AV0}$ for the model $a$. The dotted line represents the chosen initial density minimum $\rho_{min0}$ according to (\ref{107}). The dashed line is the initial density maximum $\rho_{max0}$ computed according to (\ref{70}). The solid line between the dotted and dashed lines is the chosen initial radial density profile $\rho_{LT0}$ as specified in (\ref{106}). The top solid line is the computed $\rho_{AV0}$. All values on the vertical axis are divided by $\rho_{b}$. Panel (b): The function $A$ as defined in (\ref{77}) for both models. The lower curve corresponds to the model $a$, the upper curve corresponds to the model $b$.
}
\end{figure}

\begin{figure}[]
\subfloat[]{\includegraphics[scale=0.3]{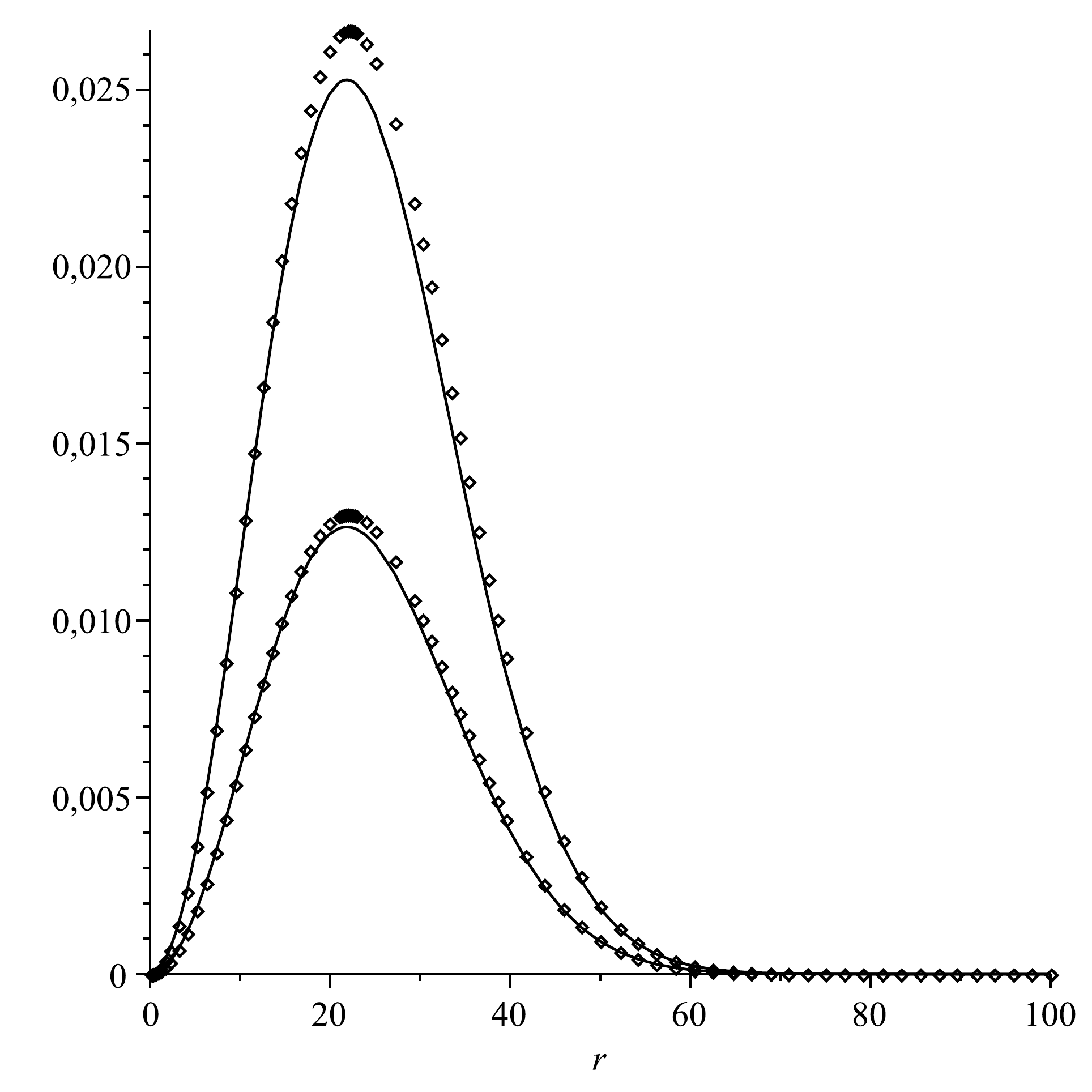}}
%\newline
\subfloat[]{\includegraphics[scale=0.3]{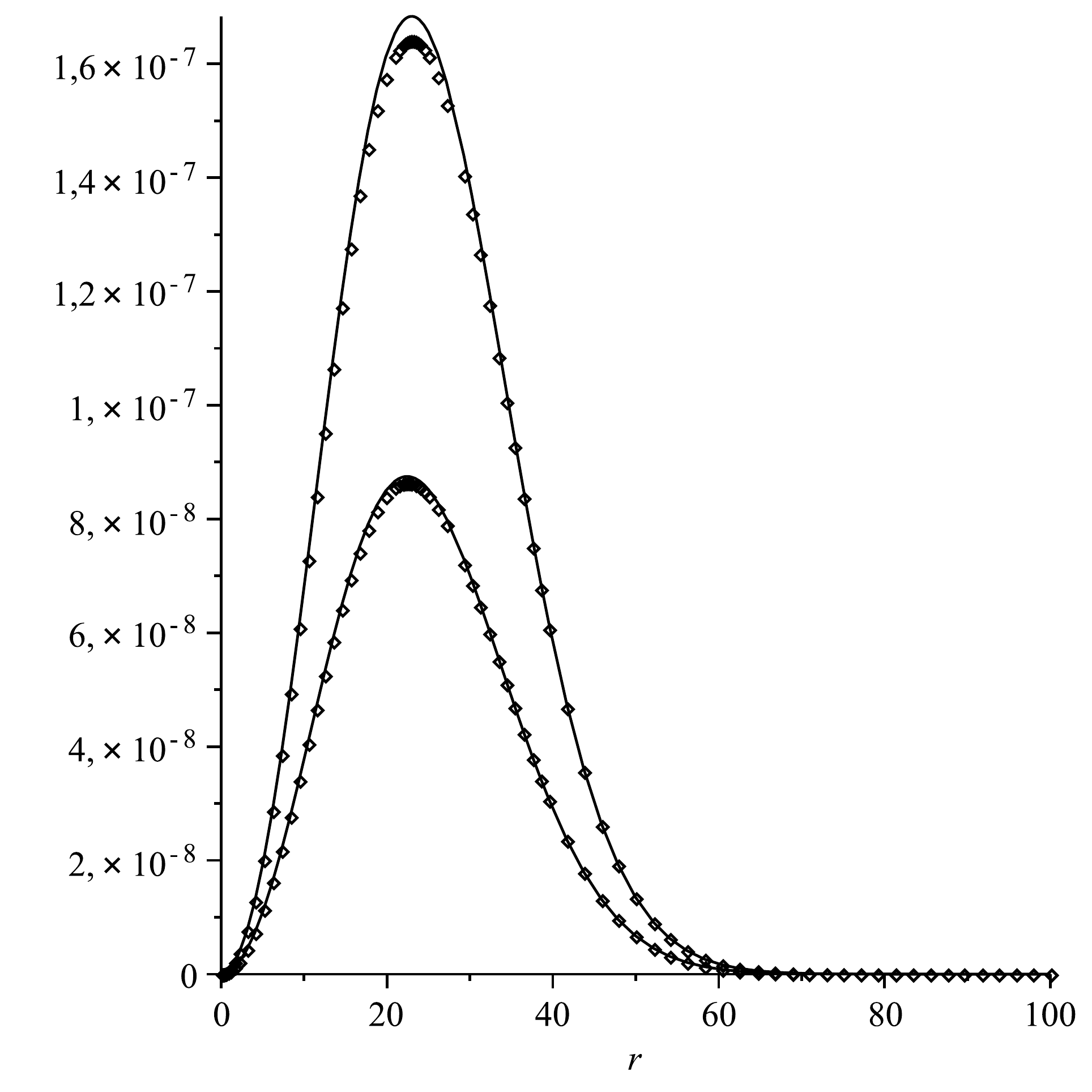}}
\caption{Panel (a): The behaviour of the initial density contrast $\frac{\Delta\rho_{0}}{\rho_{b0}}$ for both models. Panel (b): the behaviour of the final density contrast $\frac{\Delta\rho}{\rho_{b}}$ for both models. The solid curve represents the exact formula (\ref{93}). The dotted curve is the density contrast as  calculated according to the approximation formula (\ref{104}). The lower and upper curves correspond to the model $a$ and $b$ respectively. The lower curve is apparently approximated better, because the radial inhomogeneity was chosen smaller.}
\end{figure} 

\section{Conclusion}

We studied model of inhomogeneity in quasispherical Szekeres model. We set up only 3 of 5 degrees of freedom, which is sufficient for studying the evolution of the density contrast. The lack of specification of the last two degrees of freedom means that we do not have the detailed information about the density distribution, particularly we do not know the position of the extreme values on the spheres. We specify the initial radial density profile $\rho_{LT0}$ which is the value of the density around the great circle that lies in the plane that defines the dipole. Next, we choose one extreme value of the density at the initial time, either $\rho_{min0}$ or $\rho_{max0}$, and the last function that we choose is the curvature function $f$. In order to choose an appropriate value for the density extremes, we investigated the shell crossing conditions in terms of density $\rho_{max}$, $\rho_{min}$, $\rho_{LT}$ and $\rho_{AV}$. We derived conditions that $\rho_{max}$ and $\rho_{min
 }$ have to satisfy in order to avoid shell crossing and we showed that in the special case $f=0$, if the conditions are fulfilled on the initial time slice, then they will hold at any time. 

Next, we derived an analytical formula for the density contrast $\Delta\rho$ as a function of $t$ and $r$. In the special case $f=0$, we derived an approximation formula that is valid if the initial inhomogeneity is small and we showed that in this approximation the density contrast is proportional to the initial density contrast and depends on time as $t^{-3}$, so there is a decrease in magnitude during time evolution, however the shape of the function is preserved. It shows that the dynamics is very simple and close to homogeneous one for small inhomogeneity confirming the expected behaviour. In this sense one may argue that a small inhomogeneity can be successfully treated in perturbation theory and the influence of nonlinearity is negligible.

The next research will be focused on the situation when the curvature function is chosen more generally and is not equal to zero.

\begin{acknowledgements}
D. V. was supported by grant GAUK 398911 and project SVV-267301. O. S. was supported by grant GA\v{C}R 14-37086G.
\end{acknowledgements}

\end{document}